\documentstyle[11pt]{article}

\font\tenrm=cmr10
\font\tenit=cmti10
\font\elevenbf=cmbx10 scaled\magstep 1
\font\elevenrm=cmr10 scaled\magstep 1
\font\elevenit=cmti10 scaled\magstep 1

\renewenvironment{thebibliography}[1]
 { \elevenrm
   \begin{list}{\arabic{enumi}.}
    {\usecounter{enumi} \setlength{\parsep}{0pt}
     \setlength{\itemsep}{3pt} \settowidth{\labelwidth}{#1.}
     \sloppy
    }}{\end{list}}

\begin{document}
\begin{flushright}
FSU--HEP--930830\\
BNL--49462\\
August 1993
\end{flushright}
\begin{center}{
\vglue 0.6cm
{\elevenbf THE $t\bar t\gamma$ BACKGROUND TO $pp\rightarrow W\gamma + X$
AT THE SSC
\footnote{To appear in the Proceedings of the Workshop {\it ``Physics at
Current Accelerators and the Supercollider''}, Argonne National
Laboratory, June~2 --~5, 1993.}
\\}
\vglue 1.0cm
{\tenrm U. BAUR \\}
\baselineskip=13pt
{\tenit Physics Department \\}
\baselineskip=12pt
{\tenit Florida State University, Tallahassee, FL 32306\\}

\vglue 0.3cm
{\tenrm A. STANGE\\}
{\tenit Department of Physics, \\
Brookhaven National Laboratory, Upton, NY 11973\\}
\vglue 0.8cm
{\tenrm ABSTRACT}}

\end{center}

\vglue 0.1cm
{\rightskip=3pc
 \leftskip=3pc
 \tenrm\baselineskip=12pt
 \noindent
We calculate the $pp\rightarrow t\bar t\gamma+X\rightarrow W\gamma + X$
cross section at SSC energies. Approximately 40\% of the total cross section
originates from photon bremsstrahlung off the final state jet in $t\bar tj$
production. Without cuts restricting the hadronic activity, the $t\bar
t\gamma$ rate is a factor 10 (2) larger than the tree level
$W\gamma$ cross section for a top quark mass of 110~GeV (200~GeV).
Imposing a jet veto cut, the $t\bar t\gamma$ rate can be
suppressed to a level well below the $W\gamma+0$~jet signal cross section.
\vglue 0.6cm}
{\elevenbf\noindent 1. Introduction}
\vglue 0.2cm
\baselineskip=14pt
\elevenrm
One of the prime targets for experiments at present and future colliders
is the measurement of the $WW\gamma$ and $WWZ$ couplings. In the
Standard Model (SM) of electroweak interactions, these couplings are
unambiguously fixed by the non-abelian nature of the $SU(2)\times U(1)$
gauge symmetry. Experiments at the Tevatron, HERA and LEP~II are
expected to measure the three vector boson couplings at the 10~--~20\%
level at best~\cite{CUR,NLO}. High precision tests have to await the SSC
or LHC~\cite{NLO,BZ,WZ}.

A suitable process to study the $WW\gamma$ vertex at the SSC is
$W^\pm\gamma$ production~\cite{NLO,BZ}. Present studies indicate that
the background from $W+1$~jet production, where the jet is misidentified
as a photon, is under control if a large photon transverse momentum
cut~\cite{SDC} is imposed. The effects of higher order QCD corrections
can largely be compensated by imposing a jet veto~\cite{NLO}. However,
due to the very large top quark production cross section at
supercollider energies, the process $pp\rightarrow t\bar
t\gamma\rightarrow W\gamma+X$ represents a potentially dangerous background.

Here we report on a calculation of the $t\bar t\gamma$ background to
$W\gamma+X$ production at the SSC. Our calculation fully incorporates
the subsequent decay of the top quarks into a $W$ boson and a $b$-quark,
and also the $W$ decay into a fermion antifermion pair.
Besides the lowest order contributions to the associated production of a
$t\bar t$ pair and a photon, we also include photon bremsstrahlung in
$t\bar tj$ events in our calculation, using the photon fragmentation
approach. If no cuts are imposed on the hadrons in $pp\rightarrow t\bar
t\gamma\rightarrow W\gamma+X$, the $t\bar t\gamma$ production rate is found
to be approximately a factor 10 (2) larger than the lowest order
$W\gamma$ cross section for a top quark mass of $m_t=110$~GeV (200~GeV).
This considerably reduces the sensitivity of the inclusive process
$pp\rightarrow W\gamma+X$ to anomalous $WW\gamma$ couplings.
However, the $t\bar t\gamma$ background can easily be reduced to a
manageable level by requiring the photon to be isolated from the hadrons
in the event, and by imposing a jet veto, {\elevenit i.e.} by
considering the exclusive reaction $pp\rightarrow W\gamma+0$~jet.
\vglue 0.3cm
{\elevenbf\noindent 2. The Inclusive $t\bar t\gamma$ Cross Section}
\vglue 0.2cm
In our calculation of the $pp\rightarrow t\bar t\gamma$ cross section we
take into account the full set of lowest order $q\bar q\rightarrow t\bar
t\gamma\rightarrow W^+W^-b\bar b\gamma\rightarrow f_1\bar f_2f_3\bar
f_4b\bar b\gamma$ and $gg\rightarrow t\bar t\gamma\rightarrow
W^+W^-b\bar b\gamma\rightarrow f_1\bar f_2f_3\bar
f_4b\bar b\gamma$ Feynman diagrams. Graphs where the photon
is radiated from one of the $t$ or $\bar t$ decay products are not
included. The contribution from these diagrams is strongly suppressed if
a photon $p_T$ cut of $p_T(\gamma) > m_t/2$ is imposed. The top quark
and $W$ boson decays are treated in the narrow width approximation in
our calculation. Our results for $pp\rightarrow t\bar t\gamma$ agree
well with those presented in Ref.~\cite{MM}.

The bremsstrahlung contribution is calculated using the QCD $q\bar
q\rightarrow t\bar tg$, $qg\rightarrow t\bar tq$ and $gg\rightarrow
t\bar tg$ matrix elements together with the leading-logarithm
parametrization of Ref.~\cite{DO} for the photon fragmentation functions
\begin{eqnarray}
\noalign{\vskip 5pt}
z D_{\gamma /q} (z,Q^2) &=&
 F \Biggl[ {Q_q^2 (2.21-1.28z+1.29z^2)z^{0.049} \over
 1 - 1.63 \ln(1-z) } + 0.0020 (1-z)^{2.0} z^{-1.54} \Biggr] ,\quad
\\
\noalign{\vskip 5pt}
z D_{\gamma /g} (z,Q^2) &=&
{0.194\over 8} \, F \, (1-z)^{1.03} \, z^{-0.97} ,
\end{eqnarray}
where $Q_q$ is the electric charge of the quark $q$ (in units of the
proton charge $e$), $F = (\alpha / 2 \pi) \ln (Q^2 / \Lambda^2_{QCD})$,
and $z$ is the momentum fraction of the quark or gluon carried by the
photon. Since $\alpha_s(Q^2) = 12\pi/[(33-2N_F) \ln(Q^2/\Lambda^2_{QCD})]$,
these fragmentation functions are proportional to $\alpha/\alpha_s$, and
the photon bremsstrahlung contribution formally is of the same order in
$\alpha$ as the lowest order $t\bar t\gamma$ cross section.

In our subsequent analysis we focus completely on $pp\rightarrow
W^+\gamma+X$. The cross sections of the $t\bar t\gamma$ background are equal
for the $W^+\gamma+X$ and $W^-\gamma+X$ channel. The $pp\rightarrow
W^-\gamma+X$ signal rate is approximately 20\% smaller than the $pp\rightarrow
W^+\gamma+X$ cross section for the cuts specified below. Our
conclusions therefore directly apply also to the $W^-\gamma$ case.

The $W^+$ boson is assumed to decay into a
$\ell^+\nu$ final state with $\ell=e,\,\mu$. In order to simulate the
finite acceptance of detectors and to reduce fake backgrounds from jets
misidentified as photons and particles lost in the beam
pipe~\cite{SDC}, we impose
the following transverse momentum, pseudorapidity and separation cuts:
\begin{eqnarray}
\noalign{\vskip 5pt}
p_T(\ell^+) > \phantom{1}25~{\rm GeV}, & \qquad & ~~~~|\eta(\ell^+)| < 3.0,\\
p_T(\gamma) > 100~{\rm GeV}, & \qquad & ~~~~|\eta(\gamma)| < 2.5,\\
p\llap/_T > \phantom{1}50~{\rm GeV}, & \qquad & \Delta R(\ell^+,\gamma) > 0.7.
\end{eqnarray}
No cuts are imposed on the $b$-quark jets and the decay products of the
second $W$ ({\elevenit i.e.} the $W^-$) in $t\bar t\gamma$ events.
$W^-\rightarrow\tau\nu_\tau$ decays
are, for simplicity, treated like $W^-\rightarrow e\nu,\,\mu\nu$.
This ignores the contribution of the one or two additional
neutrinos in $W\rightarrow\tau\nu$ to the $p\llap/_T$ vector. Since the
$\tau$ decay channel of the $W^-$ contributes only about 11\% to the
total $W$ decay rate, the error introduced by ignoring the subsequent
$\tau$ decay is at best of the order of a few per cent, and thus much
smaller than other uncertainties in the calculation, {\elevenit e.g.}
those originating from the choice of the factorization scale $Q^2$.
For the parton distribution functions we use the HMRSB set~\cite{MRS},
evaluated at $Q^2=\hat s/4$, where $\hat s$ is the parton center of mass
energy squared.

Figure~1 shows the $p_T(\gamma)$ distribution for $pp\rightarrow t\bar
t\gamma+X\rightarrow\ell^+p\llap/_T\gamma+X$ at the SSC for
$m_t=110$~GeV (solid line), which
approximately corresponds to the present lower top quark mass
limit~\cite{LISS},
and $m_t=200$~GeV (dashed line), which is at the upper end of the
range currently believed to be consistent with the SM~\cite{SCH}.
The photon bremsstrahlung cross section is approximately
40~--~65\% of the lowest order $pp\rightarrow t\bar t\gamma$ rate over
the entire $p_T(\gamma)$ range shown in Fig.~1. The shape of the photon
transverse momentum distribution depends on the top quark mass, with
the $p_T(\gamma)$ distribution becoming harder for increasing values of
$m_t$.
\begin{figure}[t]
\vskip 9.5cm
\includegraphics{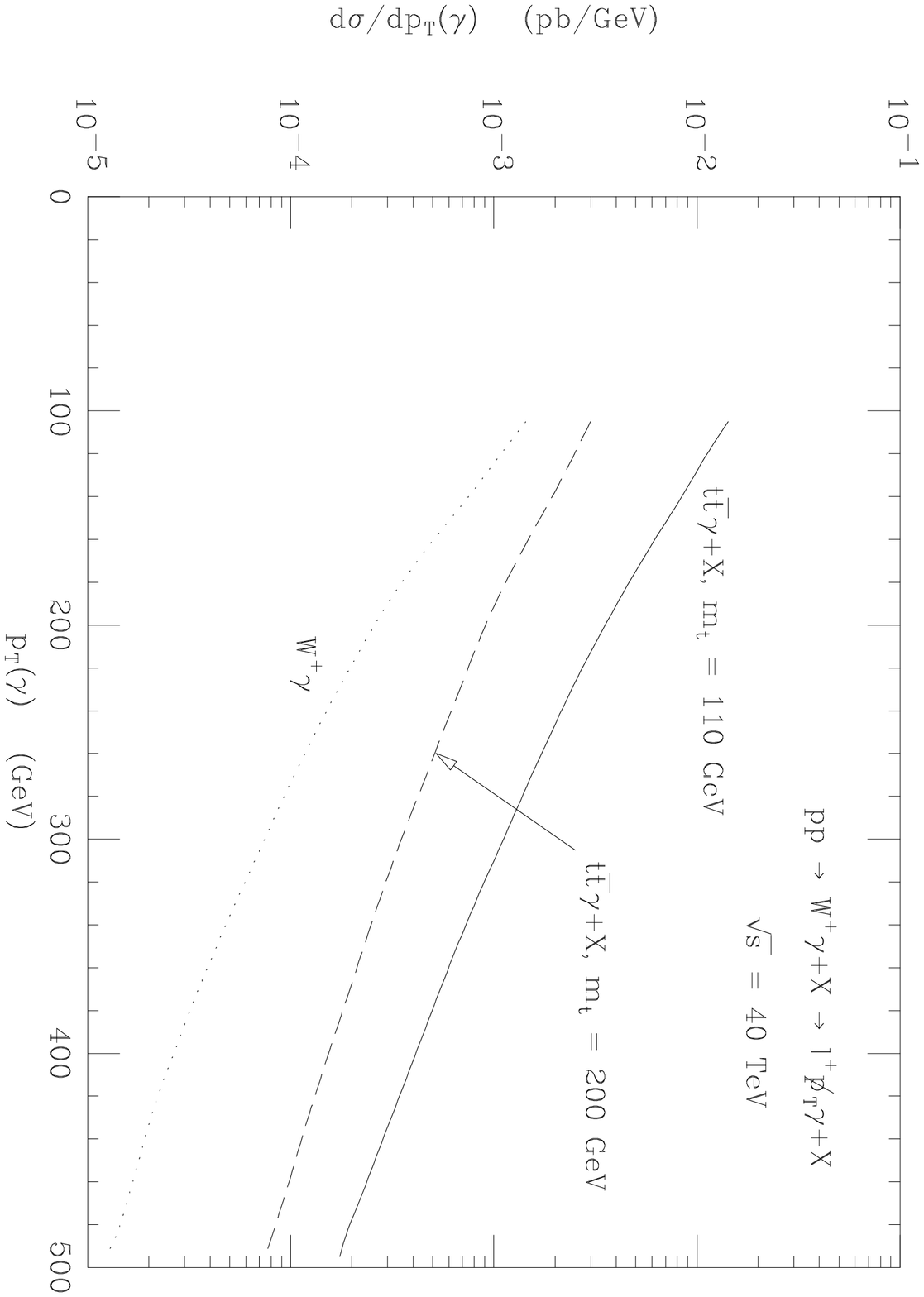}
\noindent Figure~1: The photon transverse momentum distribution for
$pp\rightarrow W^+\gamma+X\rightarrow\ell^+p\llap/_T\gamma+X$ at the
SSC. The solid (dashed) line shows the result for $t\bar t\gamma+X$
production for $m_t=110$~GeV (200~GeV). The dotted line gives the tree
level SM prediction of the $W^+\gamma$ signal. The cuts imposed are
summarized in Eqs.~(3) -- (5).
\end{figure}

The dotted curve in Fig.~1 shows the lowest order prediction of the
photon transverse momentum distribution for the $W^+\gamma$ signal. The
$t\bar t\gamma$ background is seen to be much
larger than the cross section of the signal over the entire
top quark mass range studied. For a heavy top quark, the background
is largest at high photon transverse momenta which is exactly the
region where non-standard $WW\gamma$ couplings result in large deviations
from the SM~\cite{BZ}. It is obvious from Fig.~1 that the $t\bar
t\gamma$ background will considerably reduce the sensitivity of
$pp\rightarrow W^+\gamma+X$ to non-standard $WW\gamma$ couplings.
\vglue 0.3cm
{\elevenbf\noindent 3. Jet Veto}
\vglue 0.2cm
Since the top quark decays predominantly into a $Wb$ final state, $t\bar
t\gamma$ events are characterized by a large hadronic activity which
frequently results in one or several high $p_T$ jets. If the second $W$
boson decays hadronically, up to four jets are possible.
This observation suggests that the $t\bar t\gamma$ background may be
suppressed by vetoing high $p_T$ jets. Such a ``zero jet'' requirement
has been demonstrated~\cite{NLO} to be very useful in reducing the size
of NLO QCD corrections in $pp\rightarrow W\gamma+X$ at SSC energies.
Present studies~\cite{SDC,GEM} suggest that jets with $p_T>50$~GeV can
be identified at the SSC without problems, whereas it will be difficult
to reconstruct a jet with a transverse momentum smaller than about
30~GeV. In the following we therefore require that
\begin{eqnarray}
\noalign{\vskip 5pt}
no~{\rm jets~with} \hskip 0.5cm p_T(j) > 50~{\rm GeV} \hskip 0.5cm
{\rm and} \hskip 0.5cm |\eta(j)|<3
\end{eqnarray}
are observed in $W\gamma$ events. In order to further suppress the
$t\bar t\gamma$ background, we also require the photon to be isolated
from the hadronic activity~\cite{GEM}:
\begin{eqnarray}
\noalign{\vskip 5pt}
\sum_{\Delta R<0.4} E_{\rm had} < 4~{\rm GeV},
\end{eqnarray}
with $\Delta R=[(\Delta\phi)^2+(\Delta\eta)^2]^{1/2}$. The photon
isolation cut is especially useful in suppressing the photon
bremsstrahlung contribution.

If the second $W$ in $t\bar t\gamma$ events decays hadronically, the number
of jets in $pp\rightarrow t\bar t\gamma\rightarrow W^+\gamma+X$ is
in general larger than for leptonic $W$ decays, and the jet veto is
more efficient. In order to reduce the
$t\bar t\gamma$ background for leptonic decays of the second $W$
sufficiently, we also impose a
veto on a second charged lepton in the pseudorapidity region
$|\eta|<3$. For leptons with a sufficiently large transverse momentum,
$p_T(\ell)>15$~GeV, it should be
possible to implement this requirement rather easily. For
small lepton transverse momenta, tracks in minimum bias events which are
misidentified as electrons, and decays in flight of
kaons and pions constitute a potential problem. Fortunately, most
$W$ decay leptons have a transverse momentum larger than 15~GeV. Our
results change by about 15\% if the charged lepton veto is only
implemented in the region $p_T>15$~GeV.
For $|\eta|>3$, we assume that leptons are not detected and
contribute to the missing transverse momentum in the event.

Figure~2 shows the photon $p_T$ distribution for $pp\rightarrow t\bar
t\gamma+X\rightarrow W^+\gamma+0$~jet
with $m_t=110$~GeV and the jet and lepton veto cuts described
above (solid line). The result is compared with the NLO QCD $pp\rightarrow
W^+\gamma+0$~jet result in the SM (dotted curve), and for a non-standard
$WW\gamma$ coupling (dashed line)
\begin{eqnarray}
\noalign{\vskip 5pt}
\lambda= {\lambda_0\over (1+m_{W\gamma}^2/\Lambda^2)^2}\, ,
\end{eqnarray}
with $\lambda_0=0.1$ and form factor scale $\Lambda=1$~TeV. For a
definition of $\lambda$ see {\elevenit e.g.} Refs.~\cite{CUR,NLO}. In the SM,
at tree level, $\lambda_0=0$. $m_{W\gamma}$ in Eq.~(8) is the $W\gamma$
invariant mass. The form factor nature of the anomalous coupling is
introduced to avoid violation of $S$-matrix unitarity~\cite{BZ}. The
${\cal O}(\alpha_s)$ $pp\rightarrow W^+\gamma+0$~jet cross section has been
obtained using the results of Ref.~\cite{NLO}.
\begin{figure}[t]
\vskip 9.5cm
\includegraphics{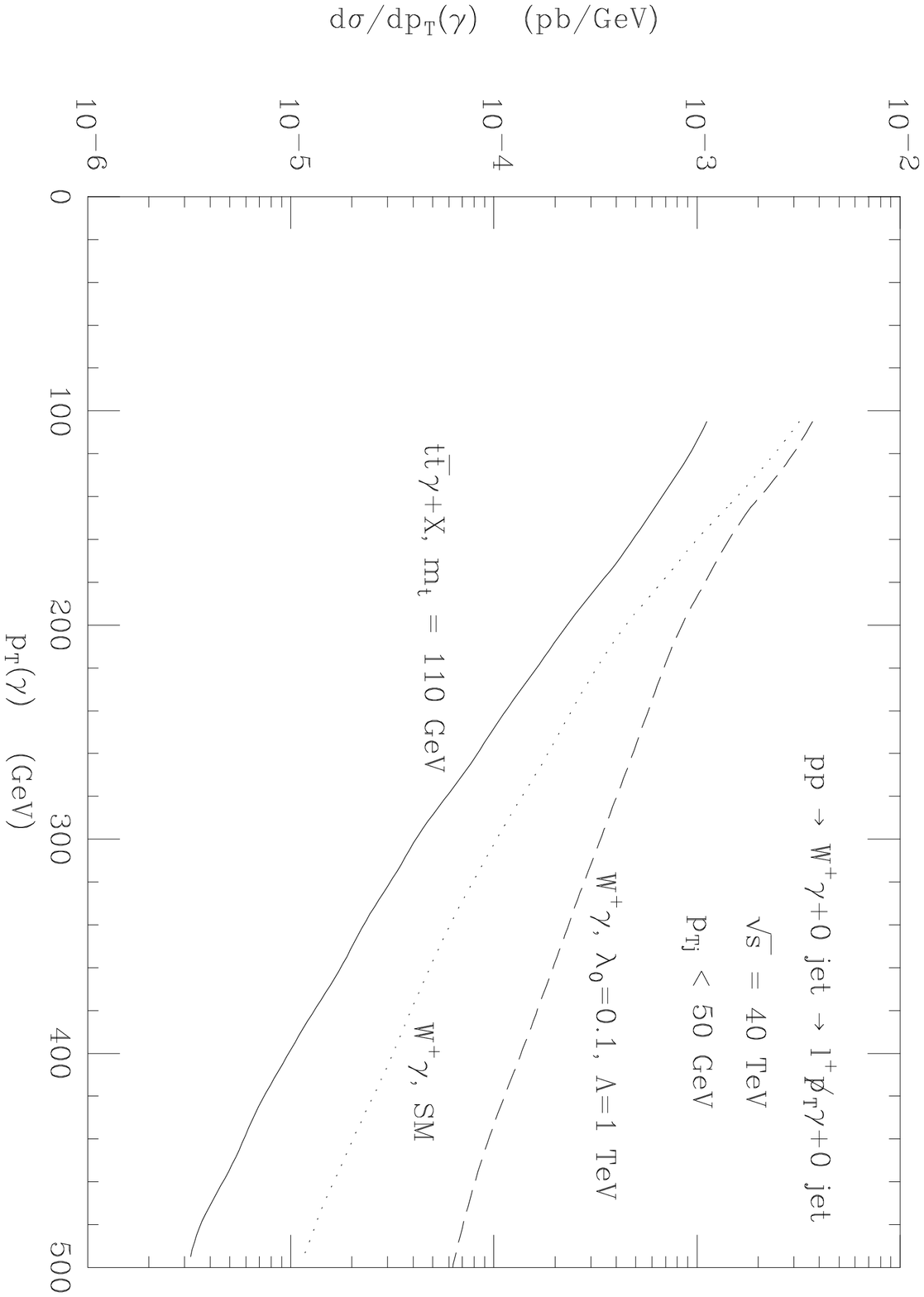}
\noindent Figure~2: The photon transverse momentum distribution for
$pp\rightarrow W^+\gamma+0~{\rm
jet}\rightarrow\ell^+p\llap/_T\gamma+0$~jet at the
SSC. The solid line shows the result for the $t\bar t\gamma$ background
with $m_t=110$~GeV. The dotted and dashed lines gives the NLO prediction of the
$p_T(\gamma)$ distribution for the $W^+\gamma+0$~jet signal in the SM,
and for a non-standard $WW\gamma$ coupling $\lambda_0=0.1$ and dipole
form factor scale $\Lambda=1$~TeV, respectively. The cuts imposed are
described in the text.
\end{figure}
NLO QCD corrections are quite significant at low photon transverse
momenta for $W\gamma$ production at
the SSC, even if a zero jet cut is imposed. We have therefore opted to
compare the $t\bar t\gamma$ background with the ${\cal O}(\alpha_s)$
prediction for the signal instead of the tree level result.

Figure~2 demonstrates that vetoing jets with $p_T(j)>50$~GeV reduces
the $t\bar t\gamma$ cross section sufficiently so that the sensitivity
to anomalous $WW\gamma$ couplings is not significantly affected. At low
(high) transverse momentum, the cuts of Eqs.~(6) and~(7) suppress the
$t\bar t\gamma$ rate by about a factor~15 (70).

Details of the various contributions to the total $pp\rightarrow
t\bar t\gamma\rightarrow W^+\gamma+0$~jet rate at the SSC for
$m_t=110$~GeV and $m_t=200$~GeV are shown in Table~1.
\begin{table}[t]
\centering
\caption
[Cross sections.]
{\elevenrm Contributions to the total $pp\rightarrow t\bar t\gamma\rightarrow
\ell^+p\llap/_T\gamma+0$~jet cross section at the SSC for two jet
defining $p_T$ thresholds. All other cuts are specified in the text.
For comparison, the last line
gives the inclusive $pp\rightarrow t\bar t\gamma+X\rightarrow
\ell^+p\llap/_T\gamma+X$ cross section, imposing only the cuts listed in
Eqs.~(3)~--~(5).
}
\vspace{6.mm}
\begin{tabular}{|c|c|c|c|}\hline
 & & \multicolumn{2}{c|}{cross section (pb)}\\[1.mm] \cline{3-4}
channel & $p_T(j)$ (GeV) & $m_t=110$~GeV & $m_t=200$~GeV \\[1.mm] \hline
direct, $W^-\rightarrow jj$ & $<50$ & 0.051 & $7.9\cdot 10^{-4}$ \\[1.mm]
 & $<35$ & 0.014 & $1.9\cdot 10^{-4}$ \\[1.mm] \hline
direct, $W^-\rightarrow\ell\nu$ & $<50$ & 0.014 & $3.5\cdot 10^{-4}$ \\[1.mm]
 & $<35$ & 0.010 & $1.8\cdot 10^{-4}$ \\[1.mm] \hline
$\gamma$ brem., $W^-\rightarrow jj$ & $<50$ & $7.7\cdot 10^{-3}$ &
$1.0\cdot 10^{-4}$ \\[1.mm]
 & $<35$ & $1.9\cdot 10^{-3}$ & $3.4\cdot 10^{-5}$ \\[1.mm] \hline
$\gamma$ brem., $W^-\rightarrow\ell\nu$ & $<50$ & $2.1\cdot 10^{-3}$ &
$5.6\cdot 10^{-5}$ \\[1.mm]
 & $<35$ & $1.4\cdot 10^{-3}$ & $3.2\cdot 10^{-5}$ \\[1.mm] \hline
total & $<50$ & 0.075 & $1.3\cdot 10^{-3}$ \\[1.mm]
 & $<35$ & 0.027 & $4.4\cdot 10^{-4}$ \\[1.mm] \hline
total, no jet veto & -- & 1.09 & 0.27\\[1.mm] \hline
\end{tabular}
\label{tab:xsect}
\end{table}
Due to the photon isolation requirement, the contribution of photon
bremsstrahlung is reduced to $\sim 15\%$ of the direct $t\bar t\gamma$
cross section. The veto on a second charged lepton in the event
significantly suppresses the channel where the second $W$ decays
leptonically.

With increasing top quark mass, the $p_T$ distribution of the $b$-quark
jets, and the jets from $W$ decay, becomes harder. The jet veto cut,
therefore, is more efficient at large values of $m_t$. For $m_t=200$~GeV the
$t\bar t\gamma+X$ cross section is reduced by about a factor~200,
whereas the rate only drops by a factor~15 for $m_t=110$~GeV (see
Table~1).

Due to the relatively large number of jets possible in $t\bar t\gamma$
events, the $pp\rightarrow t\bar t\gamma\rightarrow W^+\gamma+0$~jet
rate depends quite sensitively on the jet defining $p_T$ threshold. This
is detailed in Table~1, where we also list the cross sections if the
jet transverse momentum threshold is lowered to 35~GeV, with all other
cuts unchanged. For a jet $p_T$ threshold smaller than about 35~GeV, SSC
detectors face increasing difficulties in reconstructing
jets~\cite{SDC,GEM}. Compared to a $p_T(j)$ threshold of 50~GeV, the
$pp\rightarrow
t\bar t\gamma\rightarrow W^+\gamma+0$~jet cross section is reduced by
more than a factor~3 if the second $W$ decays hadronically. On the other
hand, for leptonic decays of the $W^-$, the rate drops only by about a
factor~1.4 to~1.8. Overall, the $t\bar t\gamma$ background can be
reduced by an additional factor~3, if a jet defining $p_T$ threshold
of 35~GeV instead of 50~GeV can be employed. The ${\cal O}(\alpha_s)$
$W\gamma+0$~jet signal cross section drops only by
approximately 20\% if the jet transverse momentum threshold is
lowered~\cite{NLO}.

The photon transverse momentum distribution of the $t\bar t\gamma$
background and the $pp\rightarrow W^+\gamma+0$~jet signal, calculated to
${\cal O}(\alpha_s)$, for a jet $p_T$ threshold of 35~GeV is shown in
Fig.~3.
\begin{figure}[t]
\vskip 9.5cm
\includegraphics{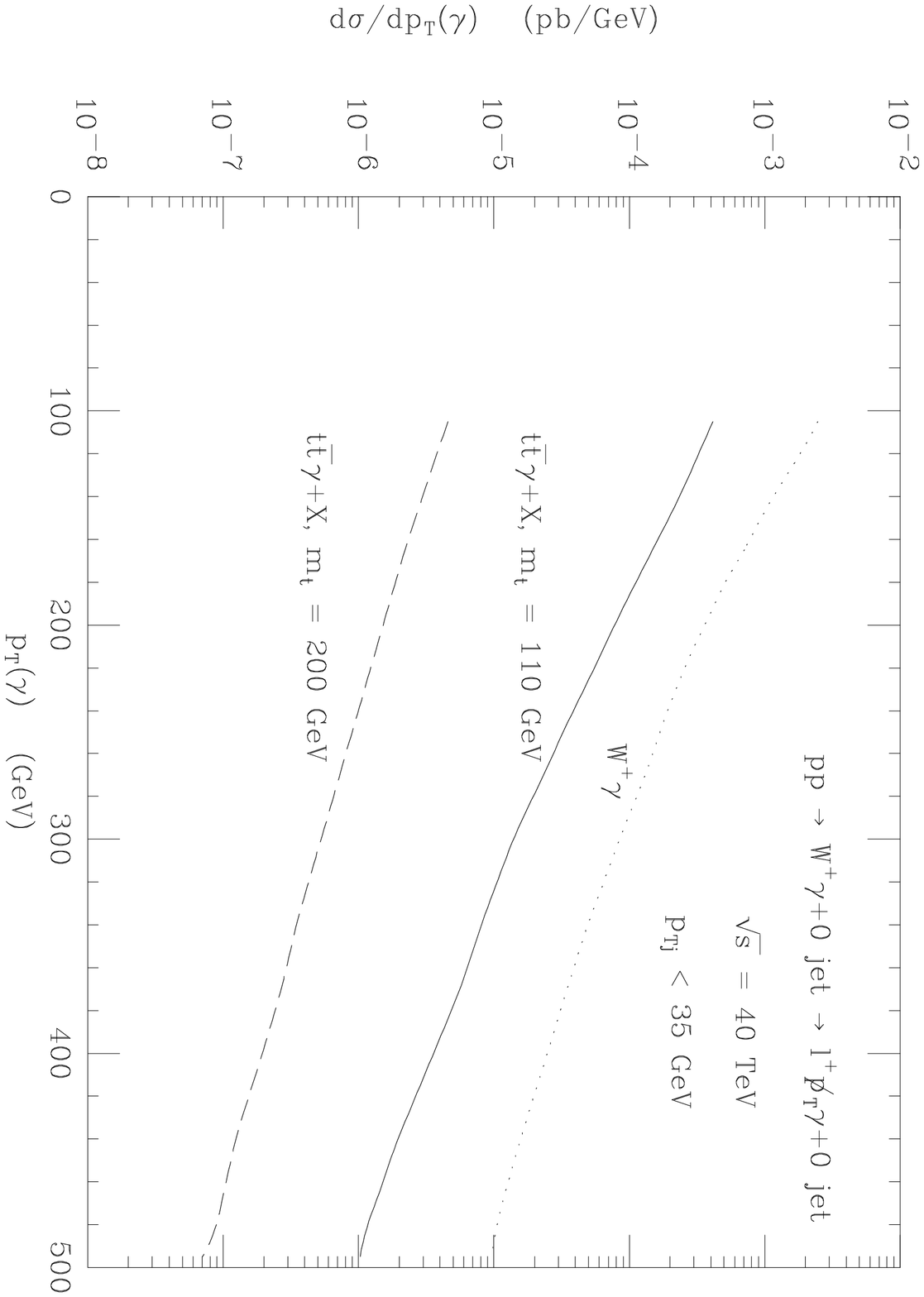}
\noindent Figure~3: The photon transverse momentum distribution for
$pp\rightarrow W^+\gamma+0~{\rm
jet}\rightarrow\ell^+p\llap/_T\gamma+0$~jet at the SSC with a jet
transverse momentum threshold of 35~GeV. The solid (dashed) line shows the
result for the $t\bar t\gamma$ background for $m_t=110$~GeV (200~GeV).
The dotted line gives the NLO prediction of the
$p_T(\gamma)$ distribution for the $W^+\gamma+0$~jet signal in the SM.
\end{figure}
The $t\bar t\gamma$ background is seen to be about one order of
magnitude below the signal (dotted line) for $m_t=110$~GeV (solid line),
and approximately two orders of magnitude for $m_t=200$~GeV (dashed
line).
\vglue 0.3cm
{\elevenbf\noindent 4. Conclusions}
\vglue 0.2cm
We have presented a calculation of $t\bar t\gamma$ production at the
SSC. Our calculation takes into account the subsequent $t\rightarrow Wb$
and $W$ decay, and incorporates the contribution originating from photon
bremsstrahlung in $t\bar tj$ events. Imposing typical photon and lepton
identification requirements, we found that the $t\bar t\gamma$
cross section is up to one order of magnitude larger than the
tree level $W\gamma$ rate. $t\bar t\gamma$ production therefore
constitutes a dangerous background to inclusive $W\gamma$ production,
$pp\rightarrow W\gamma+X$, which significantly reduces the sensitivity
of this process to anomalous $WW\gamma$ couplings.

In general, the $W\gamma$ system originating from $t\bar t\gamma$ events
is accompanied by one or several jets. Vetoing all jets with $p_T(j)
>50$~GeV and a second charged lepton
in the pseudorapidity region $|\eta(j)|<3$, the $t\bar t\gamma$
background can be suppressed to a level well below the $W\gamma+0$~jet
signal. If the jet defining transverse momentum threshold can be
reduced to 35~GeV, the $t\bar t\gamma$ background cross section is at
most 10\% of the $W\gamma$ signal rate.
\vglue 0.3cm
{\elevenbf\noindent 5. Acknowledgements}
\vglue 0.2cm
We would like to thank S.~Errede and S.~Keller for stimulating
discussions. This research was supported in part by the
U.~S.~Department of Energy under Contract No.~DE-FG05-87ER40319, and
Contract No.~DE-AC02-76CH00016.
\vglue 0.3cm
{\elevenbf\noindent 6. References}
\vglue 0.2cm

\end{document}